%% file: main.tex
\documentclass[prx,twocolumn,superscriptaddress,longbibliography]{revtex4-2}
\usepackage{bm}
\usepackage{amsmath, amsfonts}
\usepackage{amssymb}
\usepackage{times}
\usepackage{graphicx}
\usepackage[table]{xcolor}
\definecolor{lightgray}{gray}{0.9}
\usepackage{mathtools}
\usepackage{float}
\usepackage[colorlinks=true,allcolors=blue]{hyperref}
\usepackage[capitalise,nameinlink]{cleveref}
\usepackage{tikz}
\crefname{section}{Sec.}{Figs.}
\graphicspath{{figures/}}
\definecolor{ube}{rgb}{0.62, 0, 0.77}

\begin{document}

\title{Localized quasiparticles in a fluxonium with quasi-two-dimensional amorphous kinetic inductors}

%\title{Localized quasiparticle limited coherence in a quasi-two-dimensional WSi disordered superconductor fluxonium}

\input{sections/authors}
\input{sections/section0_abstract}
\maketitle 
\input{sections/section1_introduction}
\input{sections/section2_resonatormeasurements}

\input{sections/section3_fluxoniummeasurements}

\input{sections/section4_conclusion}

\input{sections/section6_acknowledgments}
%\appendix

\input{sections/sectionA1_fabrication}
\input{sections/sectionA2_modeling}
\input{sections/sectionA3_resonatorevaluation}
\input{sections/sectionA4_fridgemeasurements}
\clearpage
\bibliography{bibliography}

\end{document}

%% file: sections/authors.tex
\author{Trevyn F. Q. Larson}
\thanks{These authors contributed equally to this work.}
\affiliation{Department of Physics, University of Colorado Boulder, Boulder, CO 80309, USA}
\affiliation{National Institute of Standards and Technology, Boulder, CO 80305, USA}

\author{Sarah Garcia Jones}
\thanks{These authors contributed equally to this work.}
\affiliation{Department of Physics, University of Colorado Boulder, Boulder, CO 80309, USA}
\affiliation{Department of Electrical, Computer \& Energy Engineering, University of Colorado Boulder, Boulder, CO 80309, USA}

\author{Tam\'as Kalm\'ar}
\thanks{These authors contributed equally to this work.}
\affiliation{Department of Electrical, Computer \& Energy Engineering, University of Colorado Boulder, Boulder, CO 80309, USA}
\affiliation{Department of Physics, Institute of Physics, Budapest University of Technology and Economics, M\H uegyetem rkp.\ 3., H-1111 Budapest, Hungary}
\affiliation{MTA-BME Superconducting Nanoelectronics Momentum Research Group, M\H uegyetem rkp.\ 3., H-1111 Budapest, Hungary}

\author{Pablo Aramburu Sanchez}
\affiliation{Department of Physics, University of Colorado Boulder, Boulder, CO 80309, USA}
\affiliation{Department of Electrical, Computer \& Energy Engineering, University of Colorado Boulder, Boulder, CO 80309, USA}

\author{Sai Pavan Chitta}
\address{Department of Physics and Astronomy, Northwestern University, Evanston, IL 60208, USA}

\author{Varun Verma}
\affiliation{National Institute of Standards and Technology, Boulder, CO 80305, USA}

\author{Kristen Genter}
\affiliation{Department of Physics, University of Colorado Boulder, Boulder, CO 80309, USA}
\affiliation{National Institute of Standards and Technology, Boulder, CO 80305, USA}

\author{Katarina Cicak}
\affiliation{National Institute of Standards and Technology, Boulder, CO 80305, USA}

\author{\\Sae Woo Nam}
\affiliation{National Institute of Standards and Technology, Boulder, CO 80305, USA}

\author{Gergő Fülöp}
\affiliation{Department of Physics, Institute of Physics, Budapest University of Technology and Economics, M\H uegyetem rkp.\ 3., H-1111 Budapest, Hungary}
\affiliation{MTA-BME Superconducting Nanoelectronics Momentum Research Group, M\H uegyetem rkp.\ 3., H-1111 Budapest, Hungary}

\author{Jens Koch}
\address{Department of Physics and Astronomy, Northwestern University, Evanston, IL 60208, USA}

\author{Ray W. Simmonds}
\affiliation{National Institute of Standards and Technology, Boulder, CO 80305, USA}

\author{Andr\'as Gyenis}
\email{andras.gyenis@colorado.edu}
\affiliation{Department of Electrical, Computer \& Energy Engineering, University of Colorado Boulder, Boulder, CO 80309, USA}
\affiliation{Department of Physics, University of Colorado Boulder, Boulder, CO 80309, USA}

%% file: sections/section0_abstract.tex
\begin{abstract}
Disordered superconducting materials with high kinetic inductance are an important resource to generate nonlinearity in quantum circuits and create high-impedance environments. In thin films fabricated from these materials, the combination of disorder and the low effective dimensionality leads to increased order parameter fluctuations and enhanced kinetic inductance values.
Among the challenges of harnessing these compounds in coherent devices are their proximity to the superconductor-insulator phase transition, the presence of broken Cooper pairs, and the two-level systems located in the disordered structure. In this work, we fabricate tungsten silicide wires from quasi-two-dimensional films with one spatial dimension smaller than the superconducting coherence length and embed them into microwave resonators and fluxonium qubits, where the kinetic inductance provides the inductive part of the circuits. We study the dependence of loss on the frequency, disorder, and geometry of the device, and find that the loss increases with the level of disorder and is dominated by the localized quasiparticles trapped in the spatial variations of the superconducting gap. 
\end{abstract}

%% file: sections/section1_introduction.tex
\section{Introduction}

When charge carriers flow in a wire, their energy is stored in the magnetic field created by the current and in the inertia of the moving particles. Because the magnetic and the kinetic energies are quadratic functions of the current $I$, we can express both of these energies as $E_\mathrm{mag} = L_\mathrm{geo}I^2/2$ and $E_\mathrm{kin} = L_\mathrm{kin}I^2/2$, where the coefficients $L_\mathrm{geo}$ and $L_\mathrm{kin}$ are the geometric and kinetic inductances. While the geometry of the wire impacts both of these quantities, the material properties strongly influence the kinetic inductance. For example, in normal metals, the kinetic contribution is negligible below THz frequencies due to the frequent collisions of the charge carriers. However, in certain superconducting systems with reduced superfluid density, the kinetic inductance can surpass the geometric inductance because the kinetic inductance scales inversely with the density of Cooper pairs $n_s$. Thus, disordered materials close to the superconductor-insulator transition~\cite{sacepe_quantum_2020} can exhibit high kinetic inductance values at the cost of being more sensitive to the breaking of Cooper pairs and increased intrinsic dissipation~\cite{PhysRevLett.120.037004}.

Both geometric and kinetic inductors play important roles in various superconducting devices. The choice of inductance is based on the application: geometric inductors are easier to fabricate and tend to have higher quality factors, while kinetic inductors are more compact with less parasitic capacitance and provide more nonlinearity. For example, most low-impedance circuits, such as resonators for dispersive readout~\cite{RevModPhys.93.025005} or RF-SQUID devices~\cite{Friedman2000}, operate based on geometric inductors, but more advanced geometric inductors can also serve as elements in a variety of qubits~\cite{PRXQuantum.2.040341}. 
High kinetic inductors, on the other hand, are key elements to achieve protection against charge fluctuations in superconducting qubits~\cite{pechenezhskiy2020}, create nonlinear circuit elements~\cite{astafiev_coherent_2012}, enhance light-matter coupling~\cite{RevModPhys.91.025005}, entangle distant electron spins~\cite{borjans_resonant_2020}, miniaturize circuit components~\cite{PhysRevApplied.20.044021}, amplify signals~\cite{doi:10.1126/science.aaa8525} and engineer frequency multipliers~\cite{10.1063/5.0191743}. They are also critical materials for fundamental science applications, such as building astronomical or single-photon detectors~\cite{Zmuidzinas2012, oripov_superconducting_2023}, establishing metrological standards~\cite{shaikhaidarov_quantized_2022} and studying superconductor-insulator transitions~\cite{Mukhopadhyay2023}.

Besides metamaterials created from aluminum-oxide-based Josephson junction arrays~\cite{doi:10.1126/science.1175552, PhysRevLett.109.137002}, disordered superconducting compounds are the main avenue to achieve high kinetic inductance values using a small footprint. The most well-known examples for disordered superconductor materials are granular aluminum~\cite{maleeva_circuit_2018,grunhaupt_loss_2018, grunhaupt_granular_2019, zhang_microresonators_2019}, NbTiN~\cite{barends_contribution_2008,samkharadze_high-kinetic-inductance_2016, muller_magnetic_2022}, NbN~\cite{annunziata_tunable_2010,niepce_high_2019, PhysRevApplied.20.044021, cecile_nbn}, TiN~\cite{vissers_low_2010,shearrow_atomic_2018, leduc_titanium_2010, amin_loss_2022}, TiAlN~\cite{gao2023unraveling}, NbSi~\cite{calvo_niobium_2014}, WSi~\cite{chiles_2020,PhysRevApplied.16.044017}, InO \cite{astafiev_coherent_2012, dupre_tunable_2017, charpentier2024}, and boron-doped silicon~\cite{PhysRevApplied.17.034057}. 

Among the disordered superconductors, tungsten silicide (WSi) has been the prominent material for fabricating superconducting nanowire single-photon detectors~\cite{10.1063/1.1388868,oripov_superconducting_2023} but its potential in coherent circuits has not been explored. WSi is an amorphous disordered superconductor alloy with a critical temperature of $T_c\approx4\,$K, a superconducting gap of $\Delta_\mathrm{WSi}\approx0.6\,$meV, and a coherence length of $\xi\approx7\,$nm, depending on the stoichiometry and the thickness of the film~\cite{Kondo1992,baek_superconducting_2011,PhysRevB.94.174509}. In amorphous disordered superconductors, the lattice and long-range order are broken, unlike in crystalline disordered superconductors, where the disorder originates from the deviations in site occupancy of its constituents. Because amorphous solids require no matching to the host substrates, devices fabricated from these materials are more robust against structural defects compared to nanowires fabricated from (poly-)crystalline superconductors (such as NbTiN or NbN). The structural homogeneity and the absence of grain boundaries enable excellent yield and consistent performance across many WSi-based devices~\cite{marsili_detecting_2013}. These favorable properties are also beneficial for using WSi as a scalable linear inductor with a small size in quantum circuits.

In this work, we use quasi-two-dimensional WSi films as a platform to create microwave resonators and fluxonium qubits. We demonstrate the compatibility of WSi with current superconducting qubit fabrication technology, with performance comparable to devices relying on other disordered materials. Our results indicate that quasiparticles localized in the spatial fluctuations of the superconducting gap~\cite{PhysRevLett.117.117002} are the main limiting source of performance in this material. Furthermore, embedding WSi in fluxonium as a linear inductor is the first step towards using WSi nanowires as nonlinear circuit elements, such as weak-link Josephson elements~\cite{RevModPhys.51.101} or quantum phase slip junctions~\cite{astafiev_coherent_2012}.

%% file: sections/section2_resonatormeasurements.tex
\section{Resonator measurements}

First, to demonstrate the feasibility of using WSi in coherent devices and investigate the origin of dissipation, we fabricated distributed and lumped-element microwave resonators in a planar geometry. Our fabrication recipe closely follows the procedure developed for nanowire single-photon detectors~\cite{baek_superconducting_2011}. Starting with a solvent-cleaned c-plane sapphire wafer, we co-sputtered tungsten and silicon, and capped the films with an approximately 2\,nm-thick silicon layer to prevent aging. Throughout this study, we kept the same chemical composition of the films used for single-photon detectors (W$_{0.85}$Si$_{0.15}$) and tuned the kinetic inductance by varying the thickness of the WSi layer. The films were patterned using optical lithography and then etched using SF$_6$ reactive ion etching to create the features for the high-kinetic inductance elements of the devices.  After fabricating the WSi parts of the circuits, we used lift-off to deposit a patterned layer of either aluminum or niobium that creates the rest of the circuitry, such as the ground planes, capacitor pads, and transmission lines. We used an \textit{in-situ} RF clean to ensure the ohmic contact between the WSi and other parts of the circuits (see Supplementary Section I). Here, we focus on films with kinetic inductance of $L_K=100\,\mathrm{pH}/\square$, and 300\,pH/$\square$, corresponding to film thicknesses $h$ of around 10\,nm and 3\,nm, respectively (results on thicker films are discussed in the Supplementary Section III). Based on the relatively small film thicknesses, which are less than or comparable to the superconducting coherence length of $\xi\approx7\,$nm, these films are quasi-two-dimensional.

\begin{figure}
\includegraphics{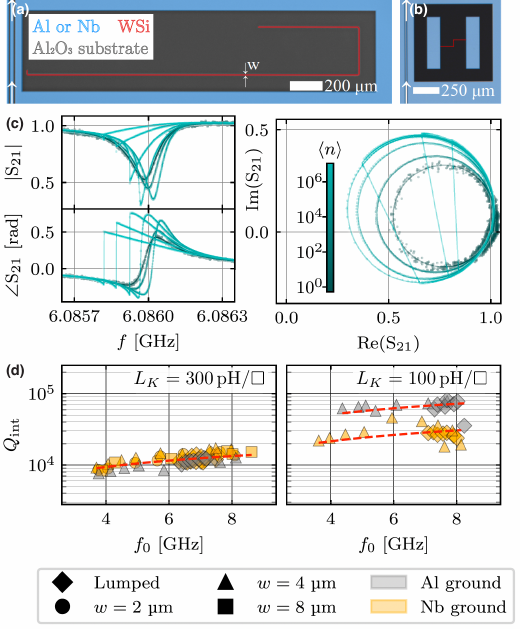}
\caption{\label{fig:fig1} (a) False-colored optical image of a WSi distributed resonator coupled to a 50-$\Omega$ transmission line. White arrows indicate the transmission line through which the $S_{21}$ scattering amplitude is measured. The width and the length of the WSi-strip are varied in the different devices. (b) Optical image of a WSi lumped element resonator. Changing the length of the WSi strips leads to different resonance frequencies. Each strip has a 4\,\textmu m width. (c) Typical amplitude ($|S_{21}|$) and phase ($\angle S_{21}$) response of a distributed resonator fabricated from the $100\,\mathrm{pH}/\square$ film with aluminum ground plane. The number of circulating photons is color-coded, and the relation between the real and imaginary parts of the scattering amplitude is shown on the right panel. Solid lines are fit to the data. (d) The quality factors below the single-photon threshold as a function of frequency of the resonators fabricated from the $L_K=300\,\mathrm{pH}/\square$ and $100\,\mathrm{pH}/\square$ films. The red dashed curves are the results of two-parameter fits, which provides the quasiparticle density ratios discussed in the main text. The performance of the $300\,\mathrm{pH}/\square$ resonators does not change between the use of aluminum or niobium for ground plane and the capacitor pads. In contrast, for devices made with $100\,\mathrm{pH}/\square$ films, devices with aluminum ground planes exhibit higher quality factors, consistent with higher dielectric loss in our liftoff niobium ground plane.}
\end{figure}

Figures \ref{fig:fig1} (a) and (b) show examples of the distributed and lumped element resonators, both of which are coupled to a 50 $\Omega$ CPW transmission line using the hanger type of arrangement~\cite{10.1063/1.2906373}. The distributed resonators contain WSi strips as the center conductors with nominal widths of $w=2, 4, 8~$\textmu m. In these distributed resonator devices, the inductance per unit length is dominated by the kinetic inductance of the wires (see Supplementary Section II for details). The lumped element resonators consist of WSi wires as the inductors and large aluminum or niobium pads as the capacitors.  We designed the resonators using finite-element electromagnetic simulators (Sonnet and HFSS) to ensure that their resonance frequencies are spread in the 4-8 GHz bandwidth of the amplifying chain. Key to our design is the vastly different geometries of the distributed and lumped element resonators, which alter the electric field participation of the WSi surface by about two orders of magnitude (see Supplementary Section II). This allows us to distinguish between intrinsic capacitive and inductive losses from coupling to parasitic two-level systems~\cite{Muller_2019} or quasiparticles~\cite{catelani_relaxation_2011}, respectively.

We extract the internal quality factors of the resonators by measuring the complex $S_{21}$ scattering parameters through the common on-chip feedlines of the devices. Figure \ref{fig:fig1} (c) reports typical measured resonance curves of one of the resonators at different photon populations. At low photon numbers, the resonance traces a circle in the complex $S_{21}$ plane as the probe frequency is swept, with an approximately symmetric amplitude response around the resonance frequency. As the power is increased, the resonance transforms into a (sweep-direction dependent) partial circle, resembling Duffing oscillator dynamics due to the nonlinearity of the kinetic inductance~\cite{Swenson_nonlinear}. The complex transmission amplitude of the resonator as a function of the probe frequency $f$ takes the form
\begin{equation}
\label{eq:s21}
    S_{21}(f) = 1 - \frac{\frac{Q_\mathrm{tot}}{Q_\mathrm{ext}} - 2 \mathrm{i}Q_\mathrm{tot} \frac{\delta f} {f_0}} {1 + 2iQ_\mathrm{tot}\frac{f-f_0}{f_0}},
\end{equation}
where $Q_\mathrm{tot}$ and $Q_\mathrm{ext}$ are the total and external quality factors of the resonator, $f_0$ is the resonance frequency, and $\delta f$ captures the asymmetry of the curves. This function is identical to the standard expression of describing the response of hanger-type resonators~\cite{10.1063/1.3692073,10.1063/5.0017378}, except here, the power-dependent resonance frequency $f_0$ is self-consistently determined for each drive frequency $f$ through a fit routine that finds the root of a cubic equation~\cite{Swenson_nonlinear,fit_routine}.  As long as the coupling is close to critical, these transmission measurements enable us to reliably extract the internal quality factors of the resonators, as $Q_\mathrm{int}^{-1} = Q_\mathrm{tot}^{-1} - Q_\mathrm{ext}^{-1}$. We achieved critical coupling in our devices by iterating over and measuring multiple variations of the designs.

Figure \ref{fig:fig1} (d) shows the extracted internal quality factors below the single-photon threshold as a function of the resonance frequency for distributed and lumped element devices fabricated from the $L_K=300\,\mathrm{pH}/\square$ and $100\,\mathrm{pH}/\square$ films. The measured quality factors range between $Q_\mathrm{int}\approx 10^4$ and $Q_\mathrm{int}\approx 10^5$, and are comparable to the internal losses of other disordered superconducting materials. In the following, we argue that the results of these measurements show multiple features that suggest that quasiparticles dominate microwave dissipation, in agreement with previous studies on granular aluminum~\cite{grunhaupt_loss_2018} and TiN~\cite{amin_loss_2022}.

\begin{figure}[h]
\includegraphics{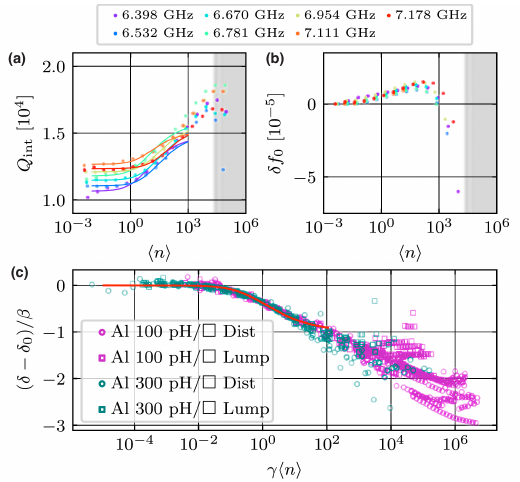}
\caption{\label{fig:fig2} (a) The internal quality factors of a set of lumped element resonators fabricated from the $L_K=300\,\mathrm{pH}/\square$ film as the number of circulating photons $\langle n\rangle$ is increased. The shaded region indicates the bifurcation regime (see Supplementary Section III). Solid lines show the result of the fit to the data using the model introduced in Ref.~\cite{grunhaupt_loss_2018}. We fit the data in the low power regime up to photon numbers equal to 1\% of the photon number corresponding to the onset of the bifurcation. (b) The relative resonance frequency change as a function of photon numbers, where $\delta f_0\langle n\rangle = \left[f_0\left(\langle n\rangle\right) - f_0\left(\langle n\rangle\approx0\right) \right] / f_0\left(\langle n\rangle\approx0\right)$. The resonance frequencies first increase with $\langle n\rangle$, signaling a reduction of the kinetic inductance. Then, close to bifurcation, the resonance frequencies suddenly drop due to the increase of the kinetic inductance. (c) Scaling of the loss tangent $\delta$ as a function of photon number for all the resonators measured on the two films, where $\beta$ and $\gamma$ are the obtained fit values using Eq.~\ref{eq:powerdependence}, and $\delta_0$ is the loss tangent at the lowest measured photon numbers. The red solid line shows the average of the fit on all the different datasets. The losses observed in all resonators can be characterized by the same photon-number-dependent behavior at low photon numbers.  At higher photon numbers, the loss tangents deviate from this behavior, which indicates a high power contribution that depends on resonator geometry and microscopic variations.}
\end{figure}

First, despite the two orders of magnitude difference in the electrical field participation ratios of the WSi surfaces in lumped element vs.~distributed resonators (Supplementary Section II), we observe no significant difference in their quality factors for the same films at given resonance frequencies. Furthermore, distributed resonators with different center strip widths show similar quality factors, while for dielectric-limited loss, a width-dependence is expected~\cite{10.1063/1.2906373}. 

Second, the quality factors of the resonators are strongly affected by the thickness of the films. In the thinner film, the phase fluctuations are enhanced and the quality factors are around $Q_\mathrm{int}\approx 10^4$. In the thicker film, we obtain quality factors closer to $Q_\mathrm{int}\approx 10^5$. Note that both films have the same composition and silicon capping layer, leading to a similar density of two-level systems in the surface oxide. Thus, the difference arises from the reduced dimensionality of the film, which can lead to increased phase fluctuations and larger quasiparticle density ratios $x_\mathrm{qp}$. 

Third, the quality factors increase monotonically with the resonance frequency across the different devices, which is in agreement with the expected loss contribution of quasiparticles~\cite{Gao2008,10.1063/1.3638063}
\begin{equation}
\frac{1}{Q_\mathrm{int}} = \frac{1}{Q_0} + \frac{\alpha}{\pi} \sqrt{\frac{2\Delta}{hf_0}}\cdot {x_\mathrm{qp}},
\label{eq:res_quality_qp}
\end{equation}
where $Q_0$ captures all losses unrelated to quasiparticles, $h$ is Planck's constant, $\Delta$ is the superconducting gap, and $\alpha$ is the kinetic inductance fraction. From the fits highlighted in Fig.~\ref{fig:fig1} (d), we extract approximate quasiparticle ratios of $x_\mathrm{qp}^{300\mathrm{pH}/\square}=3.9\cdot10^{-5}$ and  $x_\mathrm{qp}^{100\mathrm{pH}/\square}=1.2\cdot 10^{-5}$ for the two films, indicating approximately three times more quasiparticles in the thinner film.

The power dependence of the quality factors and the resonance frequencies can reveal more information about the quasiparticle dynamics in these resonators. In Fig.~\ref{fig:fig2} (a), we show a representative set of examples of how the quality factors depend on the photon number (results on the other resonators are reported in Supplementary Section III).  As the number of circulating photons is increased, the quality factors initially improve and then drop near the nonlinear bifurcation threshold.  The initial increase of the quality factors is similar to the results reported on granular aluminum~\cite{grunhaupt_loss_2018} and TiN~\cite{amin_loss_2022}, where it was argued that this characteristic is due to the recombination of localized quasiparticles. As discussed in detail in Ref.~\cite{grunhaupt_loss_2018}, central to this behavior is the presence of two species of quasiparticles: localized and mobile. In disordered superconductors, the spatial variations of the superconducting order parameter can be significant~\cite{PhysRevLett.101.157006, sacepe_localization_2011}, leading to local regions where quasiparticles can be trapped. When microwave power is increased, the localized quasiparticles can be excited and lifted out of the trapping regions of the local confining potential. This allows the unpaired quasiparticles to recombine faster, and as a result, the quality factor increases. According to the phenomenological model based on the recombination rates presented in Ref.~\cite{PhysRevLett.101.157006}, the internal quality factor evolves as
\cite{grunhaupt_loss_2018}
\begin{equation}
\frac{1}{Q_\mathrm{int}}=\frac{1}{Q_0}+\beta\left[\frac{1}{1+\frac{\gamma \langle n\rangle}{1+\frac{1}{2}(\sqrt{1+4 \gamma \langle n\rangle}-1)}}-1\right],
    \label{eq:powerdependence}
\end{equation}
where $Q_0$ describes the internal loss unrelated to quasi-particles, $\langle n \rangle$ is the average photon population, and $\beta$ and $\gamma$ describe the quasiparticle-photon coupling. We fit the power-dependence of the resonance frequency in the low photon population, below the onset of bifurcation [Fig.~\ref{fig:fig2} (a)]. The model captures the general trend, but the exact shape of the data deviates from the theoretical expectation. Upon further increase of the microwave power, which is outside the applicability of the model, the oscillating current is increased in the resonators, leading to broken Cooper pairs and a reduction in the quality factors~\cite{PhysRevLett.112.047004}. 

In addition to the behavior of the quality factors, the power dependence of the resonance frequency further supports that the number of quasiparticles first decreases with power and then increases. As Fig.~\ref{fig:fig2} (b) shows, the resonators first exhibit an increasing resonance frequency with microwave power, which indicates a decrease in inductance, and an increase in $n_s$, since  $L_\mathrm{kin}\propto 1 /n_s$. This is consistent with the reduction of the localized quasiparticle density due to their enhanced recombination after their escape from the shallow traps. At higher power, close to bifurcation, we observe a decrease in the resonance frequency, which is interpreted as the standard Kerr shift of the kinetic inductors~\cite{maleeva_circuit_2018,PhysRevApplied.20.044021}.

Finally, in Fig.~\ref{fig:fig2} (c), we demonstrate that a similar power-dependent behavior describes the other studied WSi resonators. We use a scaling method to show that in the low-photon regime, the decreased internal loss of the resonators can be explained with the same model. While the loss rates in all the measured resonators follow the universal curve at low photon numbers, the deviations of the individual resonance frequencies become significant at higher driving powers. This signals that losses in the bifurcation regime are less understood and more sensitive to the details of the resonator parameters, geometries, and materials. As a summary, our measurements on the microwave behavior of WSi resonators strongly suggest that the main loss mechanism at low driving powers is the presence of localized quasiparticles.

%% file: sections/section3_fluxoniummeasurements.tex
\section{Fluxonium measurements}

\begin{figure*}
\centering
\includegraphics[width=\textwidth]{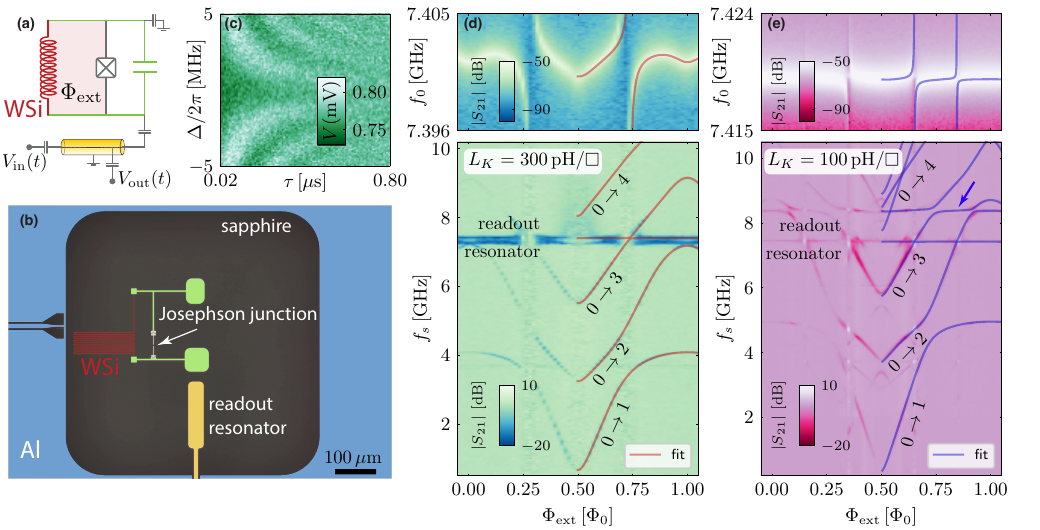}
\caption{(a) Circuit schematics of the WSi-fluxonium and (b) optical false-colored image of one of the devices. The WSi strip forms the inductor (red) for the fluxonium. The shunting capacitance (green), the ground plane (blue), and the center pin of the coplanar readout resonator (yellow) are made from an aluminum thin film. The device is biased by external magnetic flux [red shaded region in (a)], and controlled and read out through a capacitively coupled resonator. (c) Ramsey oscillation of the first excited state of the fluxonium fabricated from the 3-nm-thick, $L_K=$300 pH/$\square$ film around zero external flux ($\Delta/2\pi$ is the detuning of the microwave drive and $\tau$ is the delay between the $\pi/2$ pulses). (d) and (e) Transmission measurement of the readout resonator (top panels), and spectroscopy measurement of the fluxonium transitions (bottom panels) as a function of the external flux for the two devices. The red and blue curves show the result of a fit using a coupled resonator-qubit model. In the bottom panel of (e), the blue arrow indicates the spurious resonance mode observed in the device. The results of the fit are the following: for the $L_K=300\,$pH/$\square$ film, $E_C/h=0.88\,$GHz, $E_J/h=2.65\,$GHz, $E_L/h=0.72\,$GHz, while for the $L_K=100\,$pH/$\square$ film, $E_C/h=0.96\,$GHz, $E_J/h=3.95\,$GHz, $E_L/h=0.74\,$GHz.}
\label{fig:fig3}
\end{figure*}

Next, we test the behavior of WSi as a linear inductor in fluxonium circuits and investigate the effects of localized quasiparticles on the lifetimes of the states. Fluxonium qubits are among the most promising quantum circuits due to their strong anharmonicities~\cite{doi:10.1126/science.1175552},  long coherence times~\cite{PhysRevX.9.041041}, and a number of high-fidelity single- and two-qubit gates~\cite{PhysRevX.11.021026,PhysRevLett.129.010502}. In fluxonium [Fig.~\ref{fig:fig3}(a)], a Josephson junction is shunted by a capacitor and an inductor, leading to the Hamiltonian
\begin{equation}
    \hat{H}_0 = 4 E_C\hat{n}^2 - E_J\cos\hat\varphi +\frac{1}{2}E_L\left(\hat{\varphi} - 2\pi\frac{\Phi_\mathrm{ext}}{\Phi_0}\right)^2.
\end{equation}
Here, $E_C$, $E_J$, and $E_L$ denote the charging, the Josephson and the inductive energies, $\hat{n}$ and $\hat{\varphi}$ are the conjugate Cooper pair number and superconducting phase operators with $\left[\hat{\varphi},\hat{n}\right] = \mathrm{i}$, $\Phi_\mathrm{ext}$ is the external flux threading the loop of the device, and $\Phi_0$ is the flux quantum. Most commonly, the inductor in fluxonium is engineered from aluminum-oxide-based Josephson junction arrays, with a few exceptions, when disordered superconductors, such as NbTiN~\cite{PhysRevLett.122.010504}, TiAlN~\cite{gao2023unraveling}, granular aluminum~\cite{grunhaupt_granular_2019} wires and geometrical inductors~\cite{PRXQuantum.2.040341} serve as the inductive shunts. In this work, we rely on WSi wires as the inductors, while keeping the rest of the circuit standard (an aluminum-oxide-based single Josephson junction with aluminum capacitors and ground planes). Similar to the resonator studies discussed above, we present results on two different devices that were fabricated from WSi films with thicknesses corresponding to kinetic inductance values of $L_K\approx300\,$pH/$\square$ and $100\,$pH/$\square$.

We fabricated both WSi-fluxonium devices following the same procedure as outlined for the resonators, with the additional step of adding the single Josephson junction using standard double-angle evaporation techniques [Fig.~\ref{fig:fig3}(b)]. The WSi strips in the two devices have the same length of $l=1960\,$\textmu m, while the nominal widths are $w=2\,$\textmu m and $w=0.66\,$\textmu m, and the thicknesses are $h\approx3\,$nm and $h\approx10\,$nm for the two films, respectively. For both devices, these parameters yield the same nominal kinetic inductance of $L\approx 295\,$nH. We choose a small $E_J/E_C\approx 4$ ratio to ensure that we can probe the states of these ``light'' fluxonium qubits with a single spectroscopic tone without more involved Raman schemes~\cite{PhysRevLett.122.010504,gyenis2019}. Finally, the fluxonium qubits are capacitively coupled to a coplanar waveguide resonator that allows us to dispersively measure the energy spectrum~\cite{RevModPhys.93.025005}.

To map out the energy structure of the qubits, we perform standard spectroscopy measurements as a function of external flux. First, we measure the resonance frequency of the coupled fluxonium-resonator system by sweeping a single tone $f_0$ around the resonance frequency of the readout resonator $f_\mathrm{res}$. The top panels in Figs.~\ref{fig:fig3}(d) and (e) display the responses of the resonators of the two circuits, which exhibit avoided crossings when a qubit transition passes the resonance. The magnitudes of the vacuum Rabi splitting differ between the two samples due to the different transition matrix elements arising from the slightly different Josephson energies of the single junctions. After determining the resonance frequency at a given flux value, we apply a second tone to the circuits and measure the change in the transmission amplitude at the resonance $S_{21}(f_\mathrm{res})$ as a function of the frequency of the spectroscopy tone $f_s$. Due to the dispersive interaction between the circuit and the resonator, when the spectroscopy tone is resonant with a transition, the transmission value $S_{21}(f_\mathrm{res})$ changes as shown in the bottom panels of Figs.~\ref{fig:fig3}(d) and (e). 

We use the scQubits package~\cite{Groszkowski2021scqubitspython, Chitta_2022} to extract the transition energies, and fit the experimental results with a coupled resonator-qubit Hamiltonian
\begin{equation}
    \hat{H}_\mathrm{total} = \hat{H}_0  + \hbar\omega^0_\mathrm{res}a^\dagger a + \sum_{i,j}\hbar g_{ij}(a + a^\dagger)|i\rangle\langle j |.
\end{equation}
Here, $\omega^0_\mathrm{res} = 2\pi f_\mathrm{res}^0$ is the bare frequency of the resonator, $g_{ij}$ is the coupling energy between the qubit states $|i\rangle$ and $|j\rangle$, while $a$ and $a^\dagger$ are the annihilation and creation operators of the photons in the resonator. Note that we observe an additional mode slightly above 8\,GHz for the fluxonium made from the $L_K=100\,$pH/$\square$ film, which we take into account by adding a second bosonic mode to the fit procedure. In both devices, the theoretical model captures the transitions with high precision including the resonator and the qubit spectra. We note that in both samples, we observe smaller inductance values than the expected nominal value, possibly due to the self-resonance of the WSi wire~\cite{PhysRevLett.122.010504}. 

\begin{figure*}
\centering
\includegraphics[width=\textwidth]{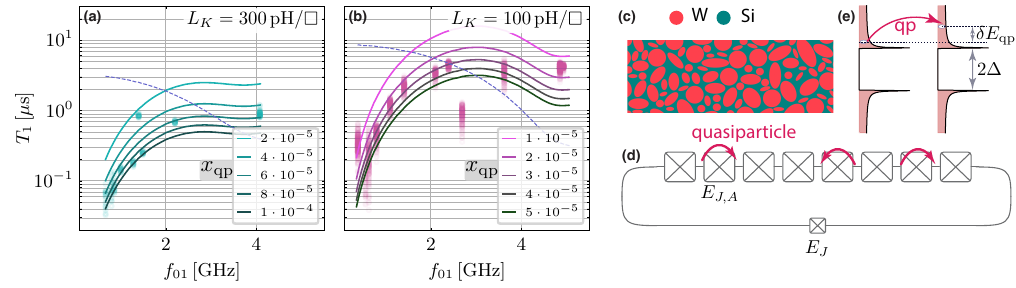}
\caption{(a) and (b) Measured relaxation $T_1$ times in log-lin scale as a function of qubit frequency $f_{01}$ for the fluxonium qubits fabricated from the $L_K=300\,\mathrm{pH}/\square$ films (cyan dots) and the $L_K=100\,\mathrm{pH}/\square$ films (pink dots). The data points at a given frequency correspond to different $T_1$ measurements acquired over several hours. The solid lines show the theoretically expected relaxation times due to quasiparticle loss at different quasiparticle density values $x_\mathrm{qp}$. The blue dashed line shows the calculated dielectric loss with $Q_\mathrm{cap} = 10^4$. The data is consistent with the predicted relaxation rates based on the quasiparticle model, suggesting an increase in the relaxation times as a function of frequency. The dielectric loss model predicts the opposite trend, where the relaxation rate decreases with increasing qubit frequency. (c) A cartoon representation of the disordered WSi film, where superconducting grains are separated by insulating barriers. (d) An approximate model of the WSi-fluxonium, where the inductor made from the disordered film is modeled as an array of Josephson junctions with the individual Josephson energies of $E_{J,A}$. Quasiparticles can tunnel through the junctions (red arrows). (e) The density of states of quasiparticles around the superconducting gap $2\Delta$ as a function of energy. Quasiparticles hopping across a junction can lead to energy loss of the qubit when the energy $\delta E_\mathrm{qp}$ is transferred to the bath.}
\label{fig:fig4}
\end{figure*}

Because the energy of the first excited state is strongly flux-tunable, these devices enable us to further study the origin of loss mechanisms associated with WSi. Using standard pulse spectroscopy and polarization saturation pulses~\cite{pop_coherent_2014}, we measure the relaxation times of the first excited states at different flux values in both devices. Figures \ref{fig:fig4}(a) and (b) show the measured $T_1$ values obtained during a period of several hours as a function of the frequency of the qubit. Similarly to the resonator measurements,  we observe faster decay rates in the fluxonium fabricated from the thinner film. Furthermore, both samples exhibit overall growing relaxation times as the qubit frequencies increase. As we show below, the frequency-dependence of the measured relaxation rates is consistent with an inductive type of loss arising from the presence of quasiparticles in the WSi inductor. Furthermore, we find that the quasiparticle densities are comparable to the ones obtained from the resonator measurements.

To understand the inductive loss in fluxonium, we revisit a model put forward to describe the macroscopic structure of granular aluminum~\cite{maleeva_circuit_2018}. According to this picture, granular aluminum consists of superconducting grains separated by insulating aluminum oxide layers, which can be approximated by a network of Josephson junctions. For the WSi films, a similar argument can be made: superconducting tungsten grains~\cite{PhysRevB.109.104519} are separated by insulating silicon barriers [Fig.~\ref{fig:fig4}(c)]. Although such a model is certainly an oversimplification of the microscopic structure of our film, it is still beneficial to obtain an approximate description of losses in WSi. We also note that even if the structural granularity is less pronounced in WSi, the strongly inhomogeneous order parameter can still lead to an emergent granularity of superconductivity~\cite{sacepe_quantum_2020}.

Assuming that we can approximate the WSi inductor as an array of Josephson junctions, we calculate the loss arising from quasiparticles tunneling across the array junctions~\cite{catelani_relaxation_2011,catelani_quasiparticle_2011} and compare the predicted loss rates with the measured relaxation data. Note that quasiparticle tunneling across the single aluminum-oxide-based junction of the fluxonium can lead to additional losses~\cite{pop_coherent_2014}, however, we find that the effect of such events is small compared to the quasiparticle tunneling events in the network of junctions in the inductor.

First, we recall the general result~\cite{catelani_relaxation_2011} that when a quasiparticle tunnels across an array junction with Josephson energy $E_{J,A}$, it can exchange energy with the qubit itself [Fig.~\ref{fig:fig4}(e)], leading to a relaxation rate of   
\begin{equation}
    \Gamma_{i\rightarrow f}^\mathrm{qp} = \left|\left\langle i \left| \sin{\frac{\hat{\theta}}{2}} \right| f \right\rangle\right | ^2 S_\mathrm{qp}(\omega).
\end{equation}
Here $\hat\theta$ is the phase drop across the junction, $\omega$ is the qubit frequency, $|i\rangle$ and $|f\rangle$ are the initial and final states of the qubit, and the normalized quasiparticle current spectral density is 
\begin{equation}
    S_\mathrm{qp}(\omega) = x_\mathrm{qp}\frac{8E_{J,A}}{\pi\hbar}\sqrt\frac{2\Delta}{\hbar\omega},
\end{equation}
where $x_\mathrm{qp}$ is the quasiparticle density ratio, and we assumed that $k_B T \ll \hbar\omega \ll \Delta$ ($k_B$ is the Boltzmann constant, $T$ is the temperature, and $\Delta$ is the superconducting gap).

In a Josephson junction array, the quasiparticle tunneling can occur across any individual junction and lead to qubit relaxation. To calculate the loss rates~\cite{catelani_relaxation_2011}, we assume that the array has $N$ identical junctions with the same Josephson energy, $E_{J,A}$, and consider the regime where the array junctions have much higher Josephson energies compared to the single junction, i.e., $E_J\ll E_{J,A}$. In this case, the array acts as a linear inductor with inductive energy of $E_L=E_{J,A} / N= \Phi_0^2/4\pi^2L$. The total transition rate in the array is the sum of the individual loss rates, such that
\begin{equation}
    \Gamma_{i\rightarrow f}^L = N\Gamma_{i\rightarrow f}^\mathrm{qp}=\left|\left\langle i \left| \hat{\varphi} \right| f \right\rangle\right | ^2\cdot S_\mathrm{qp}^L(\omega),
\label{eq:xp_loss1}
\end{equation}
where the noise spectral density is related to the total inductive energy
\begin{equation}
    S_\mathrm{qp}^L(\omega)  = x_\mathrm{qp}\frac{2E_L}{\pi\hbar}\sqrt\frac{2\Delta_\mathrm{WSi}}{\hbar\omega}.
\label{eq:xp_loss2}
\end{equation}

Interestingly, this result can be interpreted as a phenomenological inductive loss. In a quantum circuit, the inductive loss stems from the resistive part of the inductor when its total impedance has both inductive and resistive components in series. In this case, we can write the impedance of the inductor as $Z_\mathrm{ind} (\omega) = i\omega L + R(\omega)$, where $L$ is the inductance and $R(\omega)$ is the frequency-dependent resistance. For a dissipative inductor, the quality factor is the ratio of the real and imaginary components, such as $Q_\mathrm{ind}(\omega) = \omega L/R(\omega)$, and the real part of the associated admittance is $\mathrm{Re}[Y_\mathrm{ind}(\omega)]\approx 1/ \omega L Q_\mathrm{ind}(\omega)$ for $R\ll \omega L$.

When the lossy inductor is embedded in a fluxonium qubit, the phase  $\hat\varphi$ across the single junction couples to the noisy bath current $I(t)$, resulting in the coupling Hamiltonian
\begin{equation}
    \hat{H}_C = \frac{\Phi_0}{2\pi} \hat\varphi I(t).
\end{equation}
Based on Fermi's golden rule~\cite{Schoelkopf2003,smith_superconducting_2020}, this coupling gives us a relaxation rate
\begin{equation}
    \Gamma_{i\rightarrow f}^I = \frac{1}{\hbar^2}\left|\left\langle i \left| \frac{\Phi_0}{2\pi}\hat{\varphi} \right| f \right\rangle\right | ^2\cdot S_\mathrm{II}(\omega).
\end{equation}
Here the current noise spectral density in the $k_B T \ll \hbar\omega$ limit is 
\begin{equation}
    S_\mathrm{II}(\omega) = 2 \hbar\omega \mathrm{Re}\left[Y_\mathrm{ind}(\omega)\right].
\end{equation}
After expressing the admittance of the inductor, we obtain the relaxation rate due to inductive loss
\begin{equation}
     \Gamma_{i\rightarrow f}^I = \left|\left\langle i \left| \hat{\varphi} \right| f \right\rangle\right | ^2\cdot \frac{2E_L}{\hbar Q_\mathrm{ind}(\omega)}.
\label{eq:inductive_loss}
\end{equation}
By comparing Eqs.~\ref{eq:xp_loss1},~\ref{eq:xp_loss2} and~\ref{eq:inductive_loss}, we find that quasiparticle tunneling across the junction array leads to an inductive loss with a quality factor of
\begin{equation}
    \frac{1}{Q_\mathrm{ind}(\omega)} = \frac{1}{\pi} \sqrt{\frac{2\Delta_\mathrm{WSi}}{\hbar\omega}} \cdot x_\mathrm{qp}.
\end{equation}

Remarkably, this formula is identical to the quality factors of a resonator due to quasiparticle losses, when the kinetic inductance fraction $\alpha=1$ [see Eq.~\ref{eq:res_quality_qp}].

Using these results, we plot the expected relaxation rates at various quasiparticle density ratios in Figs.~\ref{fig:fig4}(a) and (b). The theoretical curves capture the trend in the measured relaxation rates in the two fluxonium qubits, with quasiparticle densities comparable to those extracted from the resonator measurements. Furthermore, this model supports that a higher quasiparticle density in the film with higher kinetic inductance correlates with lower lifetimes. 

Finally, we also investigate the effect of dielectric loss on the relaxation rates and show that the frequency dependence of this type of loss is inconsistent with the data. When modeling dielectric loss, we consider that the qubit couples through the charge operator $\hat{n}$ to the voltage $V(t)$ of a noisy bath~\cite{smith_superconducting_2020}. By introducing the capacitive quality factor $Q_\mathrm{cap}$, the relaxation rate in the $k_B T \ll \hbar\omega$ limit is
\begin{equation}
     \Gamma_{i\rightarrow f}^C = \left|\left\langle i \left| \hat{\varphi} \right| f \right\rangle\right | ^2\cdot \frac{\hbar\omega^2}{4E_C Q_\mathrm{cap}}.
\label{eq:capacitive_loss}
\end{equation}
In Figs.~\ref{fig:fig4}(a) and (b), we plot the expected relaxation rates when the capacitive quality factor is $Q_\mathrm{cap}=10^4$ with blue dashed curves. These theoretical curves fall short of predicting the observed frequency dependence of the relaxation times, as they decrease with qubit frequency, whereas the measured lifetimes increase with frequency. This further supports that the relaxation in WSi-fluxonium occurs via quasiparticles rather than due to dielectric loss, similar to the results of our resonator studies.

%% file: sections/section4_conclusion.tex
\section{Conclusions}

In this work, we investigate the behavior of microwave resonators and fluxonium devices incorporating high kinetic inductance quasi-two-dimensional WSi wires. Our combined study indicates that localized quasiparticles represent the dominant loss channels in these devices. These quasiparticles also appear to give rise to nonmonotonic nonlinearities as a function of drive strength, including a Kerr-free operation point. 

High kinetic inductance materials are an important resource for quantum circuits, but their utility is often limited by increased loss. Such increased loss has been observed across a variety of materials with significantly different elemental compositions. While several new works, including the work presented here, attribute this increased loss to quasiparticles and inductive loss channels, it remains unclear if there are fundamental limits involving tradeoffs of loss, inductance, and nonlinearity of kinetic inductance devices. Given the large parameter space of disordered and amorphous superconductors, understanding of the underlying quasiparticle dynamics is one of the critical goals for optimizing performance.

%% file: sections/section6_acknowledgments.tex
\section{Acknowledgments}
The research was sponsored by the Army Research Office under Grant Number W911NF-22-1-0050. The views and conclusions contained in this document are those of the authors and should not be interpreted as representing the official policies, either expressed or implied, of the Army Research Office or the U.S. Government. The U.S. Government is authorized to reproduce and distribute reprints for Government purposes notwithstanding any copyright notation herein. Certain commercial materials and equipment are identified in this paper to foster understanding. Such identification does not imply recommendation or endorsement by the National Institute of Standards and Technology, nor does it imply that the materials or equipment identified are necessarily the best available for the purpose.
S.G.J. acknowledges support from the National Science Foundation Graduate Research Fellowships Program under Grant No. DGE 2040434. T.F.Q.L acknowledges support in part by an appointment to the NRC Research
Associateship Program at the National Institute of Standards and Technology, administered by the Fellowships
Office of the National Academies of Sciences, Engineering, and Medicine.
G.F. acknowledges support by the János Bolyai Research Scholarship of the Hungarian Academy of Sciences, and the Ministry of Culture and Innovation and the
National Research, Development and Innovation Office
within the Quantum Information National Laboratory of
Hungary (Grant No. 2022-2.1.1-NL-2022-00004).
T.K. acknowledges support from the Europe-Colorado Program at the College of Engineering and Applied Science at the University of Colorado Boulder.

%% file: sections/sectionA1_fabrication.tex
\section*{Supplementary Section I: Fabrication}

We begin device fabrication by solvent cleaning a 3-inch c-plane sapphire wafer with two rounds of sonication in acetone and isopropanol for 2 minutes each (subsequently referred to as a solvent clean). The cleaned wafer is then blown dry with N$_2$ gas and put through a 3-minute spin, rinse, and dry (SRD) sequence in DI water.  Once dry, we deposit a WSi film by co-sputtering tungsten and silicon in an inert (argon) atmosphere with a manual AJA Sputtering Deposition System.  All three film thicknesses (3, 10, and 30 nm) have the same chemical composition (W$_{0.85}$Si$_{0.15}$), achieved with a constant chamber pressure of 1.2 mTorr during deposition. Film thickness (and consequently, kinetic inductance)  is tuned by changing the deposition time.  The required deposition times for each film thickness and corresponding kinetic inductance are available in Table \ref{tab:appendix-wsi-fab}.  To avoid oxidation, all films are capped with 2 nm of amorphous Si, sputter deposited \textit{in-situ} at a chamber pressure of 9 mTorr for 15 seconds. 

 Since our thin WSi films are mostly translucent, we pattern optical and e-beam alignment marks on the edges of the wafer with a 75 nm layer of gold by liftoff.  These alignment marks are used for all sequential lithography steps.  To begin, we spin a thin layer of resist adhesion promoter (P20), then a $\sim$ 1 \textmu m layer of SPR 660 resist that we bake at 95 $^\circ$C for 1 minute on a hot plate before patterning it in an ASML 5500/100D wafer stepper. After exposure, the wafer is post-baked at 110 $^\circ$C for 1 minute on a hot plate and developed for 30 seconds in MF26A.  Finally, the wafer is run through an SRD sequence before deposition.  The gold layer is deposited in an Angstrom electron-gun evaporator and lifted off with PG remover overnight at room temperature before it is finally rinsed with IPA and an SRD.  
 % Heating the PG remover for 10 minutes before rinsing with isopropanol can also assist in the liftoff process.
 
 The WSi wires are patterened with optical lithography and subtractive plasma etching. We use the same resist preparation and exposure procedures as in previous steps for the gold alignment marks. The WSi is patterend by SF$_6$ plasma (20 sccm SF$_6$, 25 mTorr of pressure, 70 W, -140 V DC bias) in an ion and plasma equipment reactive ion etcher (IPE RIE). Pre-etch, the chamber is conditioned with a 10-minute O$_2$ clean and a 5-minute SF$_6$ pre-condition. Etch times for each film thickness are available in Table \ref{tab:appendix-wsi-fab}.  After etching, the wafer undergoes another solvent clean to strip the remaining photoresist and is then run through an SRD once more. At this point, all WSi has been etched off the wafer except the wires themselves.

\begin{table}[]
\tiny
\resizebox{.8\linewidth}{!}{%
\begin{tabular}{c|c|c|c}
\hline
\begin{tabular}[c]{@{}c@{}}Thickness\\ {[}nm{]}\end{tabular} & \begin{tabular}[c]{@{}c@{}}Kin. Ind.\\ {[}pH / $\square${]}\end{tabular} & \begin{tabular}[c]{@{}c@{}}Deposition\\ {[}m : s{]}\end{tabular} & \begin{tabular}[c]{@{}c@{}}Etch\\ {[}m : s{]}\end{tabular} \\ \hline\hline
3     & 300     & 0 : 24     & 0 : 24     \\ \hline
10    & 100     & 1 : 18     & 0 : 54     \\ \hline
30    & 33      & 3 : 54     & 2 : 39     \\ \hline
\end{tabular}%
}
\caption{Co-sputtering deposition and reactive ion etch times for the three WSi film thicknesses fabricated in this study.}
\label{tab:appendix-wsi-fab}
\end{table}

After fabricating the WSi parts of the circuits, we deposit and liftoff either a 150-nm thick evaporated Al or a 100-nm thick sputtered Nb layer to create the rest of the circuitry such as the ground planes, capacitor pads, and feedlines as follows. First, we spin a thin layer of P20 for adhesion, followed by a 0.2 \textmu m thick layer of LOR 3A resist. We then bake the wafer on a hot plate for 5 minutes at 150 $^\circ$C. After baking, we spin a layer of SPR 660 resist with the same recipe as in the previous steps (excluding an additional P20 layer) and expose the new pattern in the ASML 5500/100D stepper once again.  Post exposure, we bake on a hotplate again for 1 minute at 110 $^\circ$C and develop using MF26A before rinsing and running the wafer through another SRD sequence.

For a Nb ground plane, we sputter a 100-nm thick film at 0.5 \AA /sec using a physical vapor deposition (PVD) system with a chamber pressure of 2.7 mTorr during deposition. For an Al ground plane, we deposit a 150 nm film at 2 \AA /sec using an Angstrom electron-gun evaporator.  In both cases, we use an \textit{in-situ} argon plasma RF clean to ensure the galvanic contact between the WSi and other parts of the circuits. We complete liftoff using the same procedure outlined above for the gold alignment marks.  This concludes the fabrication procedure for the lumped and distributed WSi resonator devices before dicing.  

For the fluxonium devices, the next step is to pattern and deposit standard double-angle-evaporated Dolan bridge Josephson junctions.  For patterning, we use electron-beam lithography with a PMGI and PMMA bilayer resist stack.  We first spin a thin layer of P20, followed by manually dispensing and spinning a 380 nm layer of PMGI and hotplate bake for 10 minutes at 200 $^\circ$C.  We then manually dispense and spin a 125-nm thick layer of PMMA and hotplate bake again for 2 minutes at 200 $^\circ$C.  We spin a second 125-nm thick layer of PMMA and hotplate bake for 10 minutes at 200 $^\circ$C for a final PMMA thickness of 250 nm.  Finally, we spin Electra 92 conductive polymer on the stack for charge dissipation during the electron-beam lithography and hotplate bake for 2 minutes at 90 $^\circ$C.  We write the junction patterns in a JEOL-6300 FS electron beam writer.

After the e-beam exposure, we ash the wafer (50 sccm O$_2$, 50 mTorr, 50 W, 20 seconds) in an IPE RIE with a 5-minute O$_2$ chamber pre-condition.  For junction deposition, we use an Angstrom electron-beam evaporator.  After loading the wafer, we first ion mill for 1 minute. Then, we complete a double-angle deposition of the junctions at $\pm 23.5 ^{\circ}$.  Between depositions, we oxidize at 4.4 Torr static oxidization for 1200 seconds to create the AlO${_x}$ barrier.

Next, we coat the wafer by photoresist and dice it before liftoff to protect the junctions from electrostatic discharge.  Once diced, we liftoff individual chips in heated PG remover at 150 $^\circ$C for 30 minutes, agitating every 10 minutes. Finally, we solvent clean, ozone, and wire-bond the chips. 

%% file: sections/sectionA2_modeling.tex
\section*{Supplementary Section II: Finite-element simulations of the resonators}

Resonators made from high kinetic inductance materials require designs tailored to both the amount of inductance of the material and the material loss. Designs for kinetic inductance resonators were verified in Ansys HFSS and Sonnet. 

\begin{figure*}
\centering
\includegraphics[width=\textwidth]{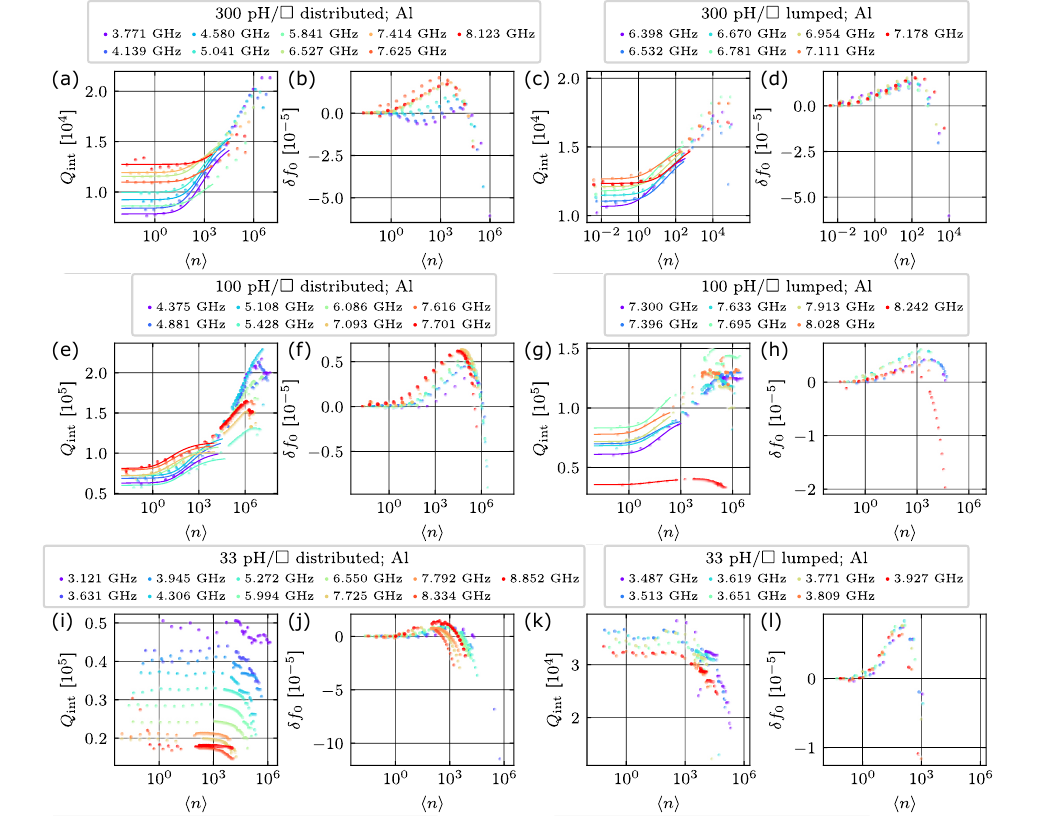}
\caption{The photon number ($\langle n\rangle$) dependent internal quality factors and resonance frequency shifts for all the tested resonators with different geometry and square inductance. For the $L_\square=300$ and $100$ pH$/\square$ films (a-h) solid lines show the fit results for the internal quality factor using Eq. \ref{eq:C2}. The observed increase in the quality factor is well captured by this formula in the low photon number regime, where bifurcation is not significant. (i-l) In the extracted quality factors of the $33$ pH$/\square$ film this initial increase is not dominant, however a clear decrease of the quality factors is visible with increasing resonance resonance frequencies.}.
\label{fig:suppA31}
\end{figure*}

Ansys simulations modeled WSi as an impedance boundary with a given reactance per square. The reactance per square was then swept, to identify a resonator frequency that is self-consistent with the given inductance per square~\cite{grunhaupt_loss_2018}. Such simulations present an effective way to estimate the inductive participation of WSi in our devices. By comparing simulations of the same design with different reactances (and thus effectively different kinetic inductances per square) we can extract frequency as a function of the kinetic inductance and find that the geometric inductance corresponds to $\approx 2.5 \mathrm{pH}/\square$. This value is in line with other standard values of geometric inductance in superconducting circuits~\cite{10.1063/1.3010859} This gives kinetic inductance participation of approximately 0.99, 0.975, and 0.93 for our $L_K=300$ $ \mathrm{pH}/\square$, $100$ $\mathrm{pH}/\square$ and $33$ $\mathrm{pH}/\square$ devices.

We also estimate the difference in the electric field participation of the lumped element and distributed designs using this method. In general, such simulations suffer from well-known issues related to nonconverging integrals~\cite{10.1063/1.3637047,10.1063/1.4934486,7745914}. In an attempt to keep such a comparison fair, we compare two simulations that show similar convergence in frequency and utilize an almost identical number of solved elements (approximately 230000).
Assuming a surface oxide thickness of approximately 2\,nm, we find that the WSi surface has a participation ratio of approximately ${p_{\text{surf}}}_{\text{WSi}} \approx 1.2 \times10^{-4}$ for the distributed designs and ${p_{\text{surf}}}_{\text{WSi}} \approx1.6 \times 10^{-6}$ for the lumped element designs, validating the intuition that the two designs have significantly different participation ratios for the WSi surface.

%% file: sections/sectionA3_resonatorevaluation.tex
\section*{Supplementary Section III: Details on resonator data analysis}
%Before fitting the resonance curves, the transmitted magnitude ($|S_{21}|$) is first converted from dBm to mW and the phase ($\angle S_{21}$) from degrees to radians. The magnitude is then normalized and the phase is unwrapped and the electrical delay is compensated for.

We use the loopfit package~\cite{fit_routine}, which is capable of handling non-linear, bifurcated resonances~\cite{Swenson_nonlinear}, to extract the resonance frequencies ($f_0$), the quality factors ($Q_{\mathrm{tot}},Q_{\mathrm{ext}}$) and the nonlinearity parameters ($a$) that characterizes the effect of bifurcation~\cite{Swenson_nonlinear}. %We utilize the $\texttt{nonlinear}$ and $\texttt{baseline}$ options of the package while excluding the mixer calibration option and setting up the sweep direction properly (increasing in our case). This step is followed by a manual check of each resonance curve and corrected by adjusting the initial guess or the fit interval when it is needed due to close lying box modes coming from the sample holder PCB\TL{I think this is too much embaressing detail}. 
In some cases, at high drive power, the resonance curves are highly distorted with a magnitude change that cannot be fitted reliably, therefore, we exclude them from further evaluation.

After extracting the resonator parameters, we determine the photon number in the resonator based on the expression of~\cite{Bruno_photon, cecile_nbn}
\begin{equation}
    \label{eq:ph_num}
    \langle n\rangle=\frac{2}{\hbar \omega_0^2} \frac{Q_{\mathrm{tot}}^2}{Q_{\mathrm{ext}}} P_{\mathrm{in}},
\end{equation}
where $P_{\mathrm{in}}$ is the microwave power incident on the resonator, $\omega_0=2\pi f_0$ is the resonance frequency, $Q_{\mathrm{ext}}$ is the external quality factor and $Q_{\mathrm{tot}}$ is the total quality factor of the resonator. We calculate $P_{\mathrm{in}}$ by correcting the VNA output power with the frequency-dependent attenuation of the input line measured at room temperature. Since the attenuation of the line decreases at low temperatures, $\langle n_{\mathrm{ph}}\rangle$ calculated with this approximation is a lower estimate.

Using the fit parameters, we determine the internal quality factor for the resonators $(Q_\mathrm{int}^{-1} = Q_\mathrm{tot}^{-1} - Q_\mathrm{ext}^{-1})$. For the $300$ and $100$ pH$/\square$ devices, $Q_\mathrm{int}$ increases with photon number, before a cut-off at the high power regime (Fig. \ref{fig:suppA31}, left panels). We interpret this behavior as the excitation and pairing of localized quasiparticles as discussed in the main text. Assuming that the evolution of the localized and mobile quasi-particle densities can be described by a Markovian process~\cite{grunhaupt_loss_2018}, for low photon numbers, the photon number dependence of the internal quality factor can be given by \cite{grunhaupt_loss_2018}
\begin{equation}
\frac{1}{Q_\mathrm{int}}=\frac{1}{Q_\mathrm{int}^0}+\beta\left[\frac{1}{1+\frac{\gamma \langle n\rangle}{1+\frac{1}{2}(\sqrt{1+4 \gamma \langle n\rangle}-1)}}-1\right],
    \label{eq:C2}
\end{equation}
where $Q_\mathrm{int}^0$ describes the internal loss unrelated to quasi-particles, $\langle n \rangle$ is the average photon population of the resonator and $\beta$ and $\gamma$ describe the quasiparticle-photon coupling approximated from the transition rates of the Markovian process. We use Eq.\,\ref{eq:C2} as a fit function on the extracted $Q_\mathrm{int}\left(\langle n\rangle\right)$ dataset. To determine the fit interval we compare the nonlinearity parameter ($a$) and exclude the points where $a$ is larger than $1\%$ of the theoretically predicted $a_{\mathrm{crit}}=4\sqrt{3}/9$ value, above which a discontinuity emerges in the resonance curves. We made this assumption to exclude the influence of high-photon number quasiparticle dynamics which is included in this phenomenological model~\cite{grunhaupt_loss_2018}. The fits are presented with solid lines in Fig.\,\ref{fig:suppA31}, and the extracted fit parameters are summarized in table \ref{tab:appendix-reson-eval}. The values for the $\beta$ and $\gamma$ parameters are on the same order of magnitude as on granular aluminum~\cite{grunhaupt_loss_2018}.

\begin{figure}
\includegraphics{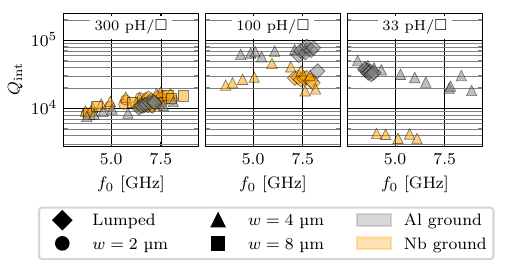}
\caption{Internal quality factors below single photon threshold as a function of device resonance frequencies for the $L_\square=300,~100$ and $33$ pH$/\square$ devices. The aluminum ground plane devices outperformed the niobium ones for the thicker films. For the $33$ pH$/\square$ devices, the quality factor decreases with increasing frequency that is caused by capacitive losses limiting the performance in these devices, rather than quasi-particle loss.}
\label{fig:suppA32}
\end{figure}

We note that the data obtained on the $33$ pH$/\square$ devices does not reproduce the above-mentioned internal quality factor increase and the captured frequency dependence of the internal loss at low photon number also shows a different trend, compared to the thinner films (Fig. \ref{fig:suppA32}.). This suggests that losses in these thicker films are due to other effects than quasiparticle loss.

The resonance frequency shift of the resonators is extracted using the same package, but for this purpose, we disable the nonlinear fitting and use it to fit an asymmetric Lorentzian curve on the $S_{21}$ datasets, up to the photon numbers where a discontinuity starts to appear as a result of bifurcation. The extracted resonance frequency shift exhibits a non-monotonic behavior as a function of photon number as discussed in the main text. In Fig.~\ref{fig:suppA31}, we report these results for all the resonators of this study. % With increasing photon number, on most resonators, the resonance frequency first increases. This is due to the reduced kinetic inductance resulting from the increased Cooper pair density, resulting from the photon-assisted pairing of localized quasiparticles. Then, with increasing further the photon number the resonance frequency decreases. This \TL{more conventional Kerr shift} is due to suppression of superconductivity at high resonator current density.

\begin{table}[]
\begin{tabular}{cccc}
\hline
\multicolumn{1}{c|}{$f_0$ {[}GHz{]}} & \multicolumn{1}{c|}{$\beta$ {[}$10^{-6}${]}} & \multicolumn{1}{c|}{$\gamma$ {[}$10^{-3}${]}} & $\delta_0$ {[}$10^{-5}${]} \\ \hline
\multicolumn{4}{c}{100 pH/$\square$ Distributed}                                                                                                                        \\ \hline
\multicolumn{1}{c|}{4.375}           & \multicolumn{1}{c|}{4.656}                   & \multicolumn{1}{c|}{79.502}                   & 1.622                      \\ \hline
\multicolumn{1}{c|}{4.881}           & \multicolumn{1}{c|}{3.966}                   & \multicolumn{1}{c|}{167.804}                  & 1.521                      \\ \hline
\multicolumn{1}{c|}{5.108}           & \multicolumn{1}{c|}{3.482}                   & \multicolumn{1}{c|}{335.690}                  & 1.466                      \\ \hline
\multicolumn{1}{c|}{5.428}           & \multicolumn{1}{c|}{4.686}                   & \multicolumn{1}{c|}{190.073}                  & 1.720                      \\ \hline
\multicolumn{1}{c|}{6.086}           & \multicolumn{1}{c|}{3.858}                   & \multicolumn{1}{c|}{289.612}                  & 1.428                      \\ \hline
\multicolumn{1}{c|}{7.093}           & \multicolumn{1}{c|}{3.559}                   & \multicolumn{1}{c|}{424.432}                  & 1.405                      \\ \hline
\multicolumn{1}{c|}{7.616}           & \multicolumn{1}{c|}{3.225}                   & \multicolumn{1}{c|}{238.005}                  & 1.273                      \\ \hline
\multicolumn{1}{c|}{7.701}           & \multicolumn{1}{c|}{3.059}                   & \multicolumn{1}{c|}{728.460}                  & 1.262                      \\ \hline
\multicolumn{4}{c}{100 pH/$\square$ Lumped}                                                                                                                             \\ \hline
\multicolumn{1}{c|}{7.300}           & \multicolumn{1}{c|}{5.641}                   & \multicolumn{1}{c|}{47.725}                   & 1.638                      \\ \hline
\multicolumn{1}{c|}{7.396}           & \multicolumn{1}{c|}{3.398}                   & \multicolumn{1}{c|}{77.397}                   & 1.432                      \\ \hline
\multicolumn{1}{c|}{7.633}           & \multicolumn{1}{c|}{4.130}                   & \multicolumn{1}{c|}{53.224}                   & 1.467                      \\ \hline
\multicolumn{1}{c|}{7.695}           & \multicolumn{1}{c|}{3.817}                   & \multicolumn{1}{c|}{38.484}                   & 1.204                      \\ \hline
\multicolumn{1}{c|}{7.913}           & \multicolumn{1}{c|}{4.433}                   & \multicolumn{1}{c|}{43.428}                   & 1.394                      \\ \hline
\multicolumn{1}{c|}{8.028}           & \multicolumn{1}{c|}{3.400}                   & \multicolumn{1}{c|}{58.591}                   & 1.277                      \\ \hline
\multicolumn{1}{c|}{8.242}           & \multicolumn{1}{c|}{3.268}                   & \multicolumn{1}{c|}{78.759}                   & 2.816                      \\ \hline
\multicolumn{4}{c}{300 pH/$\square$ Distributed}                                                                                                                       \\ \hline
\multicolumn{1}{c|}{3.771}           & \multicolumn{1}{c|}{45.353}                  & \multicolumn{1}{c|}{16.170}                   & 12.983                     \\ \hline
\multicolumn{1}{c|}{4.139}           & \multicolumn{1}{c|}{39.185}                  & \multicolumn{1}{c|}{19.458}                   & 12.090                     \\ \hline
\multicolumn{1}{c|}{4.580}           & \multicolumn{1}{c|}{34.405}                  & \multicolumn{1}{c|}{11.018}                   & 10.960                     \\ \hline
\multicolumn{1}{c|}{5.041}           & \multicolumn{1}{c|}{27.230}                  & \multicolumn{1}{c|}{10.320}                   & 10.122                     \\ \hline
\multicolumn{1}{c|}{5.841}           & \multicolumn{1}{c|}{25.930}                  & \multicolumn{1}{c|}{5.003}                    & 11.629                     \\ \hline
\multicolumn{1}{c|}{6.527}           & \multicolumn{1}{c|}{16.345}                  & \multicolumn{1}{c|}{6.709}                    & 8.688                      \\ \hline
\multicolumn{1}{c|}{7.414}           & \multicolumn{1}{c|}{9.541}                   & \multicolumn{1}{c|}{19.156}                   & 8.416                      \\ \hline
\multicolumn{1}{c|}{7.625}           & \multicolumn{1}{c|}{13.793}                  & \multicolumn{1}{c|}{4.068}                    & 9.117                      \\ \hline
\multicolumn{1}{c|}{8.123}           & \multicolumn{1}{c|}{22.841}                  & \multicolumn{1}{c|}{0.224}                    & 7.851                      \\ \hline
\multicolumn{4}{c}{300 pH/$\square$ Lumped}                                                                                                                            \\ \hline
\multicolumn{1}{c|}{6.398}           & \multicolumn{1}{c|}{28.887}                  & \multicolumn{1}{c|}{133.440}                  & 9.331                      \\ \hline
\multicolumn{1}{c|}{6.532}           & \multicolumn{1}{c|}{26.100}                  & \multicolumn{1}{c|}{48.252}                   & 9.030                      \\ \hline
\multicolumn{1}{c|}{6.670}           & \multicolumn{1}{c|}{24.779}                  & \multicolumn{1}{c|}{45.101}                   & 8.696                      \\ \hline
\multicolumn{1}{c|}{6.781}           & \multicolumn{1}{c|}{23.994}                  & \multicolumn{1}{c|}{104.948}                  & 8.441                      \\ \hline
\multicolumn{1}{c|}{6.954}           & \multicolumn{1}{c|}{21.434}                  & \multicolumn{1}{c|}{37.558}                   & 8.246                      \\ \hline
\multicolumn{1}{c|}{7.111}           & \multicolumn{1}{c|}{16.420}                  & \multicolumn{1}{c|}{51.912}                   & 7.869                      \\ \hline
\multicolumn{1}{c|}{7.178}           & \multicolumn{1}{c|}{16.903}                  & \multicolumn{1}{c|}{22.186}                   & 8.102                      \\ \hline
\end{tabular}
\caption{The extracted loss parameters using the loss model described in \cite{grunhaupt_loss_2018}.}
\label{tab:appendix-reson-eval}
\end{table}

%% file: sections/sectionA4_fridgemeasurements.tex
\section*{Supplementary Section IV: Fridge Measurement Setup}

\begin{figure}[h]
    \centering
    \includegraphics[trim={1.5cm 0.8cm 1.5cm 1cm},clip,width=\columnwidth]{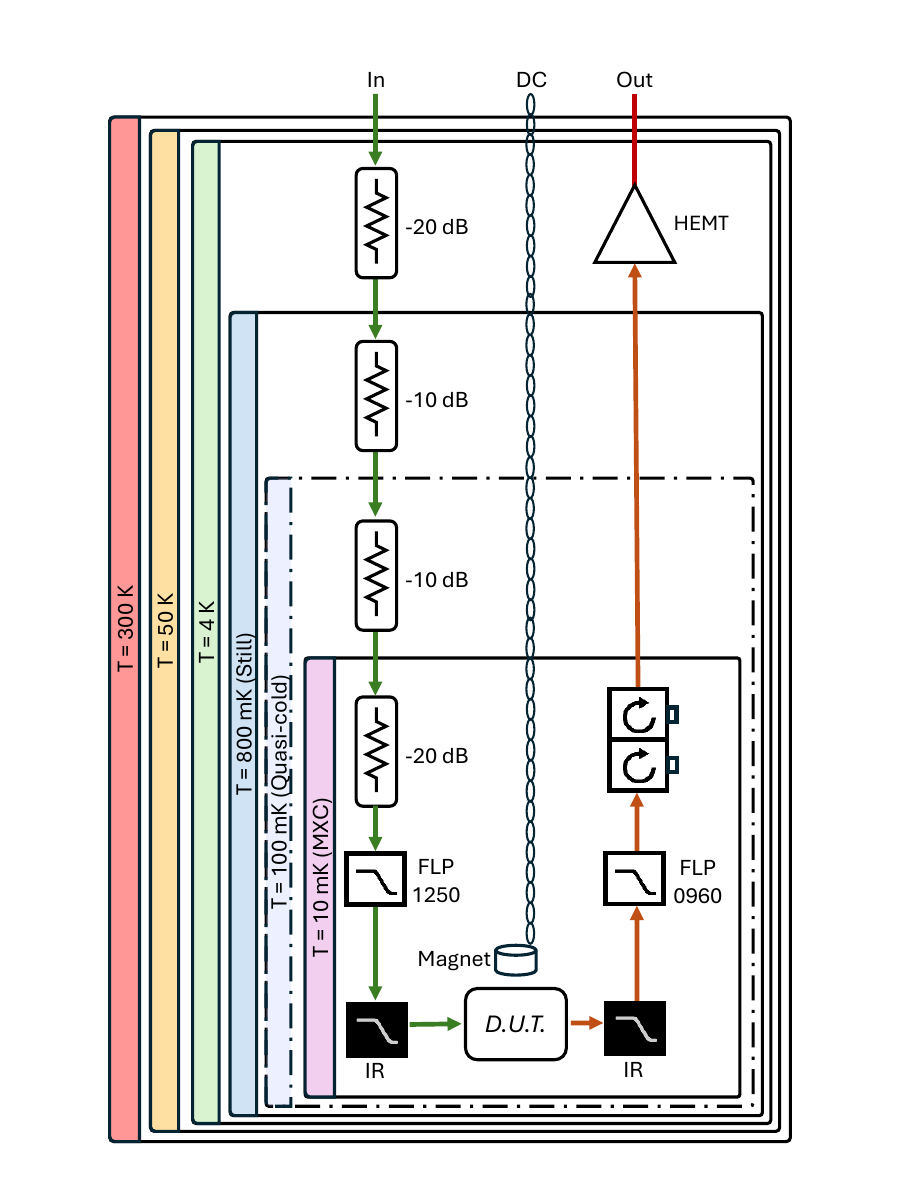}
    \caption{\label{fig:fridgewiring} Wiring diagram of the dilution fridge measurement setup.}
\end{figure}

A diagram of our fridge wiring setup is shown in Figure \ref{fig:fridgewiring}. The low pass filters on the input and output lines have cuttoff frequencies of 12.5 and 9.6 GHz, respectively, and the filters directly before and after the device under testing (D.U.T.) are both eccosorb filters for blocking infrared signals. For all measurements, the samples were housed in a light-tight aluminum shield nested inside a $\mu$-metal shield anchored to a copper cold finger on the mixing chamber flange. 

The room temperature electronics consisted of two general configurations: one for continuous-wave measurements and one for time-domain measurements. The continuous-wave measurements were taken using a vector network analyzer.  The qubit measurements were completed using the Quantum Machines OPX system.

